\documentclass[%
aps,prb,
superscriptaddress,
preprint,
rsi,%
amsmath,amssymb,
]{revtex4-1}

\usepackage{graphicx}% Include figure files
\usepackage{dcolumn}% Align table columns on decimal point
\usepackage{bm}% bold math
\usepackage{xcolor}
\usepackage{mathptmx}
\usepackage{amsthm,amsmath,amssymb}
\usepackage{mathrsfs}
\usepackage[raggedright]{titlesec}
\newcommand\rxout{\bgroup\markoverwith{\textcolor{red}{\rule[.5ex]{2pt}{.6pt}}}\ULon}

\begin{document}

\title{Deep learning enables accurate sound redistribution via nonlocal metasurfaces}
\author{Hua Ding}
\author{Xinsheng Fang}
\author{Bin Jia}
\author{Nengyin Wang}
\affiliation{Institute of Acoustics, School of Physics Science and Engineering, Tongji University, Shanghai 200092, China}
% \author{Xiaolei Ru}
% \author{Gang Yan}
% \affiliation{School of Physics Science and Engineering, Tongji University, Shanghai 200092, China}
\author{Qian Cheng}
\email{q.cheng@tongji.edu.cn}
\author{Yong Li}
\email{yongli@tongji.edu.cn}
\affiliation{Institute of Acoustics, School of Physics Science and Engineering, Tongji University, Shanghai 200092, China}

\date{\today}
\begin{abstract}

Conventional acoustic metasurfaces are constructed with gradiently ``local'' phase shift profiles provided by subunits. The local strategy implies the ignorance of the mutual coupling between subunits, which limits the efficiency of targeted sound manipulation, especially in complex environments. By taking into account the ``nonlocal'' interaction among subunits, nonlocal metasurface offers an opportunity for accurate control of sound propagation, but the requirement of the consideration of gathering coupling among all subunits, not just the nearest-neighbor coupling, greatly increases the complexity of the system and therefore hinders the explorations of functionalities of nonlocal metasurfaces. In this work, empowered by deep learning algorithms, the complex gathering coupling can be learned efficiently from the preset dataset so that the functionalities of nonlocal metasurfaces can be significantly uncovered. As an example, we demonstrate that nonlocal metasurfaces, which can redirect an incident wave into multi-channel reflections with arbitrary energy ratios, can be accurately predicted by deep learning algorithms. Compared to the theory, the relative error of the energy ratios is less than 1\%. Furthermore, experiments witness three-channel reflection with three types of energy ratios of (1, 0, 0), (1/2, 0, 1/2), and (1/3, 1/3, 1/3), proving the validity of the deep learning enabled nonlocal metasurfaces. Our work might blaze a new trail in the design of acoustic functional devices, especially for the cases containing complex wave-matter interactions.
\end{abstract}

\maketitle

\noindent The emergence of acoustic metamaterials, which are artificial materials with targeted functionalities, have significantly broaden the fields of physical and material science \cite{Liu2000Science,Fang2006NM,Yang2008PRL,Liang2012PRL,Kaina2015Nature,Cummer2016NRM}. Novel acoustic manipulations based on metamaterials have been realized, such as acoustic focusing \cite{Li2009NM,Zhu2011NP,Chen2018NC}, acoustic cloaking \cite{Popa2011PRL,Zhu2011PRL,Zhang2011PRL}, acoustic holography \cite{Marzo2015NC,Melde2016Nature,Zhu2018NC}, and asymmetric acoustic transmission \cite{Liang2009PRL,Liang2010NM,Fleury2014Science,Devaux2015PRL}. In the last decade, acoustic metasurfaces, as 2D acoustic metamaterials, stand out as distinct choices to manipulate sound for their vanishing thickness \cite{Li2013SR,Zhao2013SR,Tang2014SR,Li2014PRApplied,Xie2014NC,Mei2014NJP,Ma2014NM,Li2015PRApplied,Li2017PRL,Zhu2017PRX,Li2018NC,Assouar2018NRM,Fu2020SA}. By employing ``locally'' designed phase shift provided by subunits of metasurfaces, fascinating acoustic manipulations are extensively explored recently, including acoustic cloaking \cite{Jin2019PRApplied}, acoustic holography \cite{Zhu2018NC}, and compact sound absorbers \cite{Mei2012NC,Ma2014NM,Yang2017MH,Huang2020SB}. However, in the most of previous designs, the subunits are designed individually according to the discretized phase shift profile.  In this way, the mutual coupling among the subunits which also contributes the wave-field forming is neglected. It therefore turns out that the designed metasurfaces suffer from the lower efficiency, especially in the cases where incident/reflected/transmitted angles are relatively large. Furthermore, to reshape wave field with fine resolution, a large number of subunits are necessary which inevitably introduce undesired loss due to the viscou-thermal effect induced by small apertures in small subunits. Recently, the concept of ``nonlocal'' metasurface is introduced to overcome these limitations \cite{Ra2017PRL,Asadchy2017PRX,Hou2019PRApplied,Ni2019PRB}. In these works, the nonlocal coupling effects among all subunits are taken into accounted to collect the contribution of nonlocality to the wave-field forming and therefore give rise to more accurate wave field manipulations. However, it also makes the nonlocal system quite complex and the relationship between desired wave field and the geometrical parameters unclear. In this situation, optimization methods are generally employed to inversely design the functionality of nonlocal metasurface. However, for different wave manipulation aims, the objective function needs to be revised correspondingly to search for the best value in the entire parameter space, resulting in the drawbacks of time-wasting and computational resource-consuming.

The development of artificial intelligence algorithms provides new insights into the inverse design of nonlocal metasurfaces. In view of its extreme success in domains correlated to the detection of diseases \cite{Ribli2018Detecting,Naseer2019Refining}, speech recognition \cite{Nassif2019Speech}, computer vision \cite{Voulodimos2018Deep}, and natural language processing \cite{Young2018Recent,Wu2019Deep}, deep learning has aroused wide attention. The excellent learning and inference capabilities of deep learning enable it to achieve the similar effect of logical calculations with physical formulas. Many works about combining deep learning with physics have been attempted, including optimization of photonic structures \cite{Liu2018Training,Sajedian2019Optimisation,Asano2018Optimization,Long2019Inverse,Tahersima2019Deep,Pilozzi2018Machine,John2018Nanophotonic}, high-resolution imaging
\cite{Liu2018Learning,Orazbayev2020Far}, and acoustic source imaging \cite{Xu2021Acoustic}. 

In this work, by harnessing the power of deep learning, we explore the capabilities of reshaping acoustic fields accurately by nonlocal metasurfaces. The complex nonlocal coupling can be learned by deep learning efficiently from the preset datasets. Then for a target wave field, deep learning can predict the structural parameters of a nonlocal metasurface in a few seconds. In the following, we will demonstrate that nonlocal metasurfaces predicted by deep learning can redirect an incident wave into multi-channel reflections with arbitrary energy ratios. Compared to the direct calculations, the relative error of the energy ratios is less than 1\%. In addition, we also perform experiments to witness the ability of the deep learning enabled nonlocal metasurfaces. 

% Inspired by this, we contrive a transient reactor via deep learning in this work, which can achieve forward or inverse prediction in a few seconds. Specifically, we will instantly obtain the corresponding diffraction coefficient ratio by inputting the structural parameters into this reactor. More important is structural parameters could be instantly obtained by inputting diffraction coefficient ratio, which not only greatly reduces the design time for researchers, but also provides the possibility of independent design for engineers who have no theoretical knowledge on acoustics. All the work in this article is based on three-channel splitters via acoustic metasurfaces which could arbitrarily distribute the energy in each channel. We firstly introduce the non-local diffraction grating principles for the structure and establish the training data set. Then, we construct a transient reactor by training prepared dataset. Finally, we verify the feasibility and efficiency of the reactor by showing predicted results and performing experiments on the 3D-printed samples.

\begin{figure}
\includegraphics[width=12.6cm]{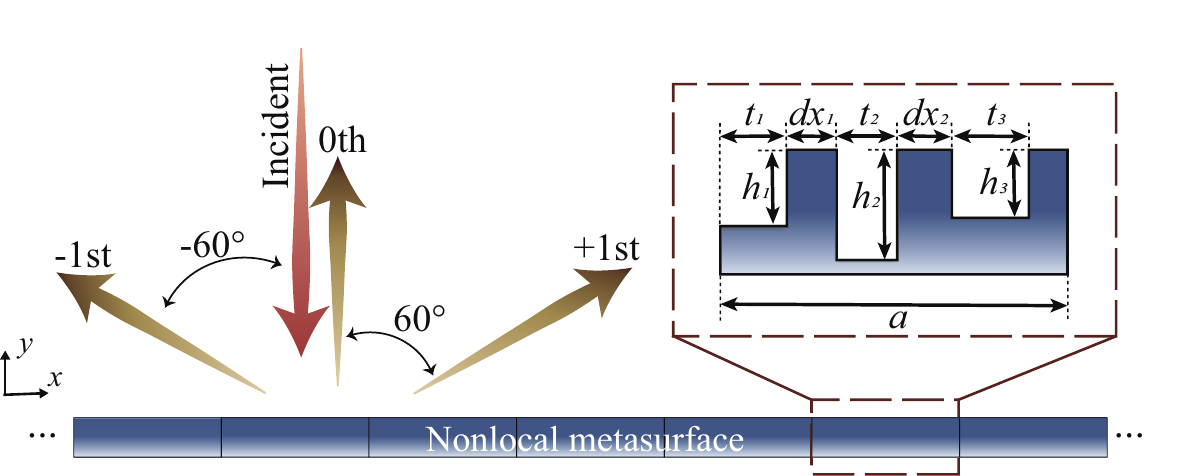}
\caption{\label{fig:1} The schematic diagram of nonlocal acoustic metasurface. The metasurface is composed of repeated supercell along $x$ direction with a period of $a$. Three grooves are etched in each supercell. ${t_l}$ and ${h_l}$ represent the width and the depth of the $l$th groove, and ${dx_l}$  refers to the distance between adjacent grooves ($l=1,\ 2,\ 3$). The incident wave along the $-y$ direction could be reflected to three allowed diffracted beams.}
\end{figure}

\noindent \textbf{Results}\\
\noindent \textbf{Theory of nonlocal metasurface.} Our proposed nonlocal metasurface is schematically shown in Fig.~\ref{fig:1}, which consists of a suitable engineered arrangement of grooves. In this work, we set 3 rectangular-shaped grooves in each period to demonstrate the ability of energy redistribution among three allowed diffraction beams. We will show that empowered by machine learning algorithms, the energy proportion of the three diffracted beams can be redistributed arbitrarily by tuning the depth (${h_l}$) and the width (${t_l}$) of each groove and the distance between two adjacent grooves (${dx_l}$). 

When an incident acoustic plane wave along with ${-y}$ direction impinges on the metasurface (i.e., incident angle  ${\theta _i}$ = $0^\circ$), the periodicity of the metasurface will give rise to a set of different diffraction modes. According to the coupled mode theory, the pressure field $p(x,y)$ and normal velocity field $v(x,y)$ in the region above the structure (Region I) can be expressed by \cite{Ni2019PRB}
\begin{equation}
{p_{\rm I}}(x,y) = A_0^+{e^{j{k_0}y}} + \sum\limits_n {A_n^-} {e^{ - j{k_{x,n}} \cdot x}}{e^{ - j{k_{y,n}} \cdot y}}, \label{eq:1}
\end{equation}
\begin{equation}
{v_{\rm I}}(x,y) =  - \frac{{A_0^+}}{{\rho \omega }}{k_0} \cdot {e^{ - j{k_0}y}} + \sum\limits_n {\frac{{A_n^ - }}{{\rho \omega }}} {k_{y,n}} \cdot {e^{ - j{k_{x,n}} \cdot x}}{e^{ - j{k_{y,n}} \cdot y}}, \label{eq:2}
\end{equation}
where ${A_0^+}$, ${A_n^-}$ are the amplitude of the incident wave and the $n$th-order of the reflected diffraction beams. ${k_{x,n}} = {k_0}\sin {\theta _i} + {G_n},$ and ${k_{y,n}} = \sqrt {{k_0}^2 - {k_{x,n}^2}},$ are the $x$ and $y$ component of the $n$th-order diffraction wave vector with ${k_0}$ being the wave number of incident waves and ${G_n} = n{2\pi/a}$ representing the reciprocal lattice vector. $\rho$ and $\omega$ refer to the mass density of the medium and the angular frequency. From the above formulas, the period $a$ of the metasurface related to $k_{x,n}$ decides the propagation characteristics of different diffraction modes. It means that by selecting a suitable $a$, the numbers of the allowed propagating diffraction orders can be tuned. 

To illustrate the nonlocal coupling effect among etched grooves, we need first get the pressure and velocity field inside each grooves. Considering the fact that the bottom of the grooves are acoustically hard where the normal velocity should be zero, the $p(x,y)$ and $v(x,y)$ in grooves (Region II) could be represented as
\begin{equation}
{p_{{\rm I}{\rm I}}}^l(x,y) = \sum\limits_k {{a_{kl}}} \cos [{\alpha _{kl}}(x - {x_l})]({e^{j{\beta _{kl}}\cdot y}} + {e^{ - j{\beta _{kl}}(y + 2{h_l})}}), \label{eq:3}
\end{equation}
\begin{equation}
{v_{{\rm I}{\rm I}}}^l(x,y) = \sum\limits_k { - \frac{{{a_{kl}}}}{{\rho \omega }}} {\beta _{kl}}\cos [{\alpha _{kl}}(x - {x_l})]({e^{j{\beta _{kl}}\cdot y}} - {e^{ - j{\beta _{kl}}(y + 2{h_l})}})
\label{eq:4}
\end{equation}
where ${a_{kl}}$ represents the amplitude of the $k$th-order of the waveguide mode, ${x_l}$ is the beginning $x$ coordination of the $l$th groove, ${\alpha _{kl}} = k\pi /{t_l}$ and ${\beta _{kl}} = \sqrt {{k_0}^2 - {\alpha _{kl}}^2}$ are the $x$ and $y$ components of the wave vector for the $k$th-order mode in the groove, respectively. Conforming to the conditions at the air-metasurface interface, sound pressure and volume velocity should be continuous at $y$ = 0. By plugging Eqs.~(\ref{eq:1})-(\ref{eq:4}) into continuum conditions and integrating the equations with orthogonal relationship, we can get the following equations:
\begin{equation}
\left( {\begin{array}{c}
  {P_1^1}  \\ 
  {P_1^2}  \\ 
  {P_1^3}  \\ 

 \end{array}} \right)A_0^ +  + \left( {\begin{array}{c}
  {P_2^1}  \\ 
  {P_2^2}  \\ 
  {P_2^3}  \\ 

 \end{array}} \right)A_n^ -  = \left( {\begin{array}{ccc}
  {P_3^1} & 0 & 0  \\ 
  0 & {P_3^2} & 0  \\ 
  0 & 0 & {P_3^3}  \\ 

 \end{array}} \right)\left( {\begin{array}{c}
  {a_{kl}^1}  \\ 
  {a_{kl}^2}  \\ 
  {a_{kl}^3}  \\ 

 \end{array}} \right),
\label{eq:5}
\end{equation}
and
\begin{equation}
{V_1}A_0^ +  + {V_2}A_n^ -  = \left( {\begin{array}{ccc}
  {V_3^1} & {V_3^2} & {V_3^3}  \\ 

 \end{array} } \right)\left( {\begin{array}{c}
  {a_{kl}^1}  \\ 
  {a_{kl}^2}  \\ 
  {a_{kl}^3}  \\ 

 \end{array} } \right),
\label{eq:6}
\end{equation}
where
\begin{equation}
P_1^l(k) = \frac{1}{{{t_l}}}\int_{{x_l}}^{{x_l} + {t_l}} {\cos {\alpha _{kl}}(x - {x_l})} dx,
\label{eq:7}
\end{equation}
\begin{equation}
P_2^l(k,n) = \frac{1}{{{t_l}}}\int_{{x_l}}^{{x_l} + {t_l}} {{e^{ - j{k_{x,n}}x}}} \cos {\alpha _{kl}}(x - {x_l})dx,
\label{eq:8}
\end{equation}
\begin{equation}
P_3^l({k_1},{k_2}) = (1 + {e^{ - j2{\beta _{{k_1}l}}{h_l}}})\frac{1}{{{t_l}}}\int_{{x_l}}^{{x_l} + {t_l}} {\cos {\alpha _{{k_1}l}}(x - {x_l})\cos {\alpha _{{k_2}l}}(x - {x_l})} dx,
\label{eq:9}
\end{equation}
\begin{equation}
{V_1}(m) =  - \frac{1}{a}\int_0^a {{k_{y,0}}{e^{j{k_{^{x,m}}}x}}} dx,
\label{eq:10}
\end{equation}
\begin{equation}
{V_2}(m,n) = \frac{1}{a}\int_0^a {{k_{y,n}}{e^{ - j({k_{^{x,n}}} - {k_{^{x,m}}})x}}} dx,
\label{eq:11}
\end{equation}
and
\begin{equation}
V_3^l(m,k) =  - (1 + {e^{ - j2{\beta _{kl}}{h_l}}})\frac{{{\beta _{kl}}}}{a}\int_{{x_l}}^{{x_l} + {t_l}} {\cos {\alpha _{kl}}(x - {x_l}){e^{j{k_{x,m}}x}}} dx,
\label{eq:12}
\end{equation}
with $m, n = 0, \pm 1, \cdots , \pm (N - 1)/2$ and $k, k_1, k_2 = 1, \cdots , K$. ${P_1^l}$, ${P_2^l}$, ${P_3^l}$ are respectively related to the the sound pressure of incident wave, reflected wave in Region I and the sound wave in Region II. And their dimensions are known as $LK$×1, $LK$×$N$ and $LK$×$LK$. $L$ is the total number of grooves in a period, $K$ is the total number of modes inside the metasurface, and $N$ is the total number of diffraction modes. In fact, the superscript and subscript of $V$ have the similar meaning as those of $P$, but $V$ is related to the volume velocity of sound waves. The dimensions of ${V_1}$, ${V_2}$, ${V_3^l}$ are $N$×1, $N$×$N$, $N$×$LK$ respectively. We simplify the above equations with matrixes and rewrite them as:
\begin{equation}
\left( {\begin{array}{c}
  {{P_1}}  \\ 
  {{V_1}}  \\ 

 \end{array} } \right)A_0^ +  = \left( {\begin{array}{cc}
  { - {P_2}} & {{P_3}}  \\ 
  { - {V_2}} & {{V_3}}  \\ 

 \end{array} } \right)\left( {\begin{array}{c}
  {A_n^ - }  \\ 
  {{a_{kl}}}  \\ 

 \end{array} } \right),
\label{eq:13}
\end{equation}
From Eq.~(\ref{eq:13}), the reflected energy of each diffraction order could be calculated. In order to show energy distribution intuitively, we utilize $I = {\left| {\frac{{A_n^ - }}{{A_0^ + }}} \right|^2}\cos {\theta _r}$ to denote the power flow ratio of each mode. Here, we choose -1st, 0th and +1st order diffractive components as reflected channels, and the corresponding diffraction angles for them are ${\theta _r}$ = $-60^\circ$, $0^\circ$, $60^\circ$. The working frequency is set to $f_0= 8000$Hz. In real situations, the harmonic modes of the metasurface and its internal modes are very complicated, so we choose $K$=20, $N$=51, $L$=3 to be as close to the actual situations as possible. Obviously, the consideration of non-locality makes the dimension of the matrixes large which means that the computational complexity is higher. This brings more troubles to inverse design.

\noindent \textbf{Deep learning model.} The special operating mechanism of deep learning will help to explore the relationship between metasurface and its response to acoustic waves. Here, we establish a tool that can achieve rapid forward and inverse design with the support of deep learning. This part of content mainly explains the process of building the transient reactor through deep learning algorithms (the specific training mechanism can be found in the Method). 
\begin{figure}
\includegraphics[width=10.6cm]{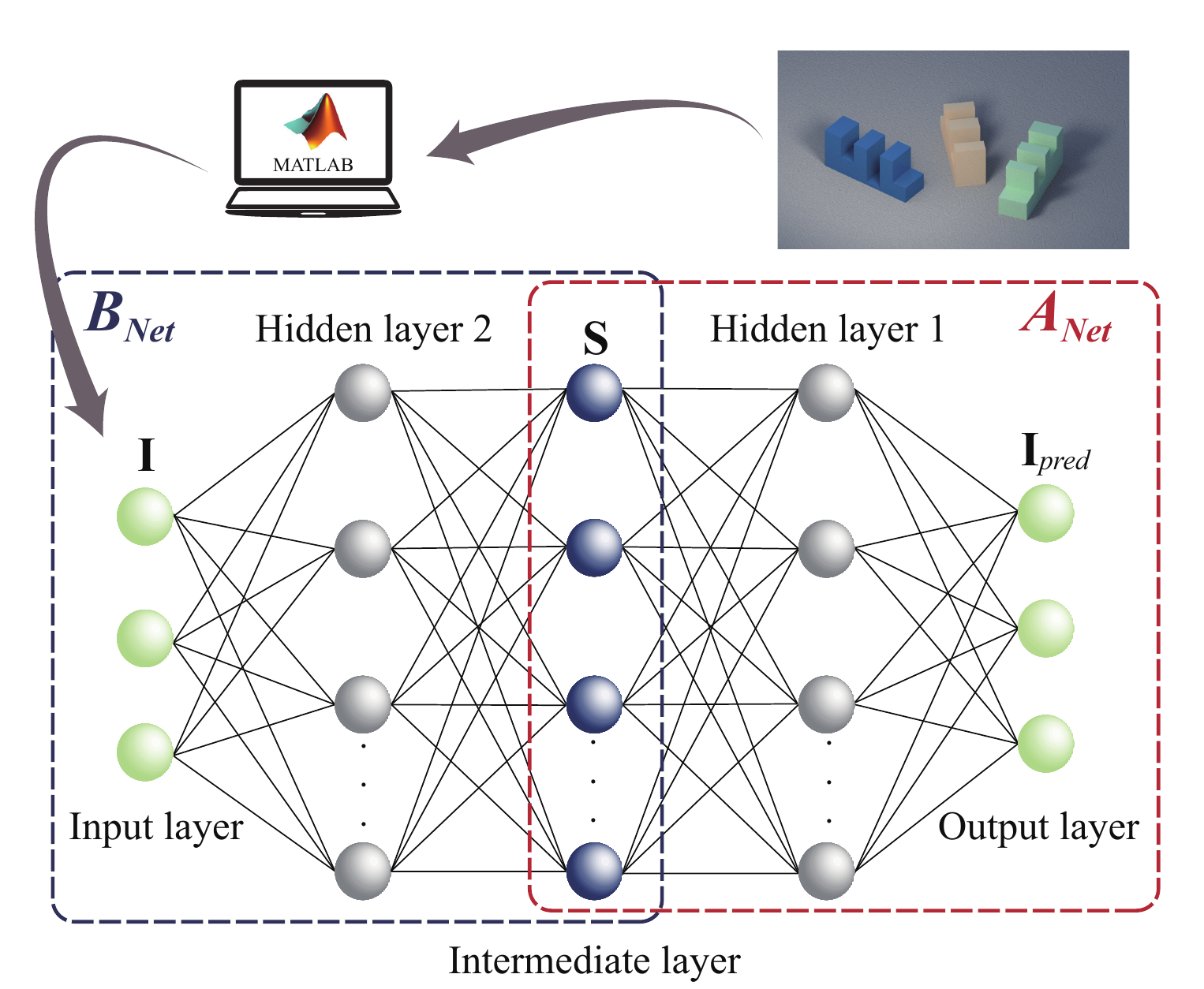}
\caption{\label{fig:2}  Schematic representation of data acquisition and the tandem network. This network is composed of pre-trained forward network ${A_{net}}$ (in the red dotted box) and inverse network ${B_{net}}$ (in the blue dotted box). The intermediate layer where the blue nodes are located is the output layer of the inverse design.}
\end{figure}

To realize the intelligent design of metasurfaces, we firstly should establish a data set. We introduce a vector
${\bf{S}} = {({h_1},{h_2},{h_3},{t_1},{t_2},{t_3},d{x_1},d{x_2})^{\text{T}}}$ which is composed of the structure parameters of metasurfaces. Meanwhile, reflected power flow ratio in three channels is expressed as a state vector ${\bf{I}} = {({I_1},{I_2},{I_3})^{\text{T}}}$, where ${I_1}$, ${I_2}$, ${I_3}$ respectively correspond to power flow of $-1$st, $0$th and $+1$st order diffractive components.The dataset based on diffraction theory for metasurfaces contains many arrays \{$\bf{S}$, $\bf{I}$\}, including 270,000 training samples and 30000 test samples. Specifically, as shown in Fig. \ref{fig:2}, we randomly generate many different sets of $\bf{S}$ and calculate their corresponding $\bf{I}$ by MATLAB. Here, we construct two network models (${A_{net}}$ and ${B_{net}}$), one of which takes $\bf{S}$ as input data and $\bf{I}$ as label data while the other is the opposite. Obviously, there is a correspondence between these two vectors, $\bf{I}$ = ${A_{net}}$($\bf{S}$), $\bf{S}$ = ${B_{net}}$($\bf{I}$), where ${A_{net}}$ is the forward deep learning network and ${B_{net}}$ is the inverse neural network. It should be noted that ${B_{net}}$ can hardly converge when it is trained in a similar way as ${A_{net}}$ since the same power-flow distribution could be accounted for different arrangement of meta-units.

To tackle this multi-value problem, we exploit tandem neural network introduced in Ref.\cite{Liu2018Training}. As Fig. \ref{fig:2} shows, the tandem network model is composed of ${A_{net}}$ and ${B_{net}}$, which connects the trained ${A_{net}}$ served as a backend behind the ${B_{net}}$. This connected network is similar to an autoencoder network (consists of two networks: encoder and decoder) in terms of structures, in which the input data in the first layer has the same physical meaning as the data in the final layer and structural parameters exist as the encoded feature data in the intermediate layer. The entire network trained successfully is actually a transient reactor. In the construction of it, the forward network with dataset \{$\bf{S}$, $\bf{I}$\} is supposed to be trained as the pre-trained network firstly. Then, when training the connected model, $\bf{I}$ is both the input data and the target value that needs to be compared with the output data. Different from the general autoencoder training mechanism which is often used for unsupervised learning, we have a prepared dataset including label data, so we can train the ``decoder'' (${A_{net}}$) first. Training part of the network in advance transforms the surjective relation between $\bf{S}$ and $\bf{I}$ into bijective relation, and in this way, the root cause of the failure to converge inverse network is resolved.

For all training processes, Adaptive Moment Estimation (Adam) optimizer \cite{Kingma2014Adam} is chosen as the global optimizer.  Forward training is done by minimizing the cost function, ${C_1} = \frac{1}{N}\sum\limits_{} {|{{\bf{I}}_{true}}}  - {{\bf{I}}_{pred}}|$, in which $N$ is the batch size of training dataset, ${{\bf{I}}_{pred}}$ is the possible power flow ratio predicted by ${A_{net}}$ and ${{\bf{I}}_{true}}$ is the target value in dataset. The value of cost function is inclined to be steady after 1500 epochs of forward training. The successful training of ${A_{net}}$ informs that deep learning model could acquire the relationship between physical quantities.  During the secondary training procedure, every layer in pre-trained forward model is frozen, that is, weights and bias of each node are not updated in layers of ${A_{net}}$ but these values in ${B_{net}}$ are optimized all the time. The cost function of this combination model can be expressed by ${C_2} = \frac{1}{N}\sum\limits_{} {|{{\bf{I}}_{true}} - {A_{net}}({\bf{S}})} |$. In a way, the ``decoder'' converts the multi-value problem into a single-value problem without complicated data filtering. As the corresponding relation between the design $\bf{S}$ and the response $\bf{I}$ changes, the training of connected network converges easily (nearly after 1500 epochs).

With the entire model trained successfully, the transient reactor for three-channel metasurfaces is completely constructed. When we take the middle layer as input and the last layer as the output, it could realize the forward data prediction. While we use the first layer as the input and the middle layer as the output, the desired structure parameters would be predicted by the middle layer, i.e., the output layer of inverse network, as long as we input the target power flow distribution. Actually, what we are concerned more about is the function implemented by the latter, because nonlocal metasurfaces do not have a clear 
inverse theoretical relationship between structures and their responses, but the realization of the model allows this theoretical relationship to be simulated. Moreover, it avoids the troubles of conventional optimization methods that require to iterate many times to search for the best value and easily fall into local optimality.

\noindent \textbf{Nonlocal metasurfaces predicted by deep leanring.} The target structures of acoustic metasurfaces could be predicted quickly and intelligently by the transient reactor. We randomly select several groups of power flow distributions with different ratios to verify the accuracy and efficiency of this reactor. Part of the prediction results is shown in Table \ref{tab:table1}. The values on the left indicate the target ratios fed to the model and the theoretical values in middle represent the real energy distributions calculated by the predicted geometric parameters as shown in the right column. All the ratios agree well with each other, which reveals
the accuracy of this reactor. 

\begin{table}
\centering
\caption{\label{tab:table1}Part of design results by deep learning model.}
\begin{ruledtabular}
\begin{tabular}{ccc}
$\bf{Target\quad value}$&$\bf{Theoretical\quad value}$&$\bf{Structural\quad parameters(unit: mm)}$\\
\hline
$[0.70,0.00,0.30]$&$[0.7006,0.0026,0.2967]$&$[9.01,22.53,10.75,8.46,7.79,8.41,8.40,8.40]$ \\
$[0.80,0.00,0.20]$&$[0.8005,0.0007,0.1988]$&$[8.71,23.22,11.32,8.42,7.62,8.25,8.21,8.25]$ \\
$[0.90,0.10,0.00]$&$[0.9000,0.0996,0.0004]$&$[8.06,23.66,14.06,8.20,6.40,7.09,7.70,8.36]$ \\
$[0.30,0.30,0.40]$&$[0.2989,0.3007,0.4004]$&$[11.80,20.09,11.07,8.39,7.93,8.38,8.47,8.47]$ \\
$[0.20,0.30,0.50]$&$[0.1990,0.2999,0.5011]$&$[12.78,20.04,10.39,8.40,7.92,8.40,8.49,8.48]$ \\
$[0.20,0.50,0.30]$&$[0.2000,0.5005,0.2995]$&$[12.83,19.46,11.79,8.43,7.56,8.49,8.49,8.50]$ \\
$[0.40,0.20,0.40]$&$[0.3998,0.2005,0.3997]$&$[11.06,20.44,11.04,8.47,8.22,8.39,8.49,8.44]$ \\
\end{tabular}
\end{ruledtabular}
\end{table}

For further verification, we choose three sets of predicted data for simulation and experiment. These three target ratios are set to be ${{\bf{I}}_{\rm I}} = [1,0,0]$, ${{\bf{I}}_{{\rm I}{\rm I}}} = [\frac{{\text{1}}}{{\text{2}}},0,\frac{{\text{1}}}{{\text{2}}}]$, ${{\bf{I}}_{{\rm I}{\rm I}{\rm I}}} = [\frac{{\text{1}}}{{\text{3}}},\frac{{\text{1}}}{{\text{3}}},\frac{{\text{1}}}{{\text{3}}}]$. For these cases, the structures designed by the transient reactor are shown as follows (unit: mm):
$${{\bf{S}}_{\rm I}} = [6.80, 23.52, 11.74, 8.19, 6.41, 7.06, 8.09, 8.19],$$
$${{\bf{S}}_{{\rm I}{\rm I}}} = [10.10, 23.01,10.12, 8.30, 7.99, 8.46, 8.37, 8.37],$$
$${{\bf{S}}_{{\rm I}{\rm I}{\rm I}}} = [11.53, 19.96, 11.53, 8.44, 7.80, 8.44, 8.46, 8.47].$$
Corresponding power flow ratios in theory are ${{\bf{I'}}_{{\rm I}}} = [0.9910,0.0026,0.0063]$, ${{\bf{I'}}_{{{\rm I}{\rm I}}}} = [0.4986,\\0.0017,0.4998]$, ${{\bf{I'}}_{{{\rm I}{\rm I}{\rm I}}}} = [0.3341,0.3315,0.3344]$, which fit well with the target values. The finite element analysis is used here for numerical calculation, which could reveal intuitive sound field distribution based on various geometric sizes.

\begin{figure}
\includegraphics[width=16.0cm]{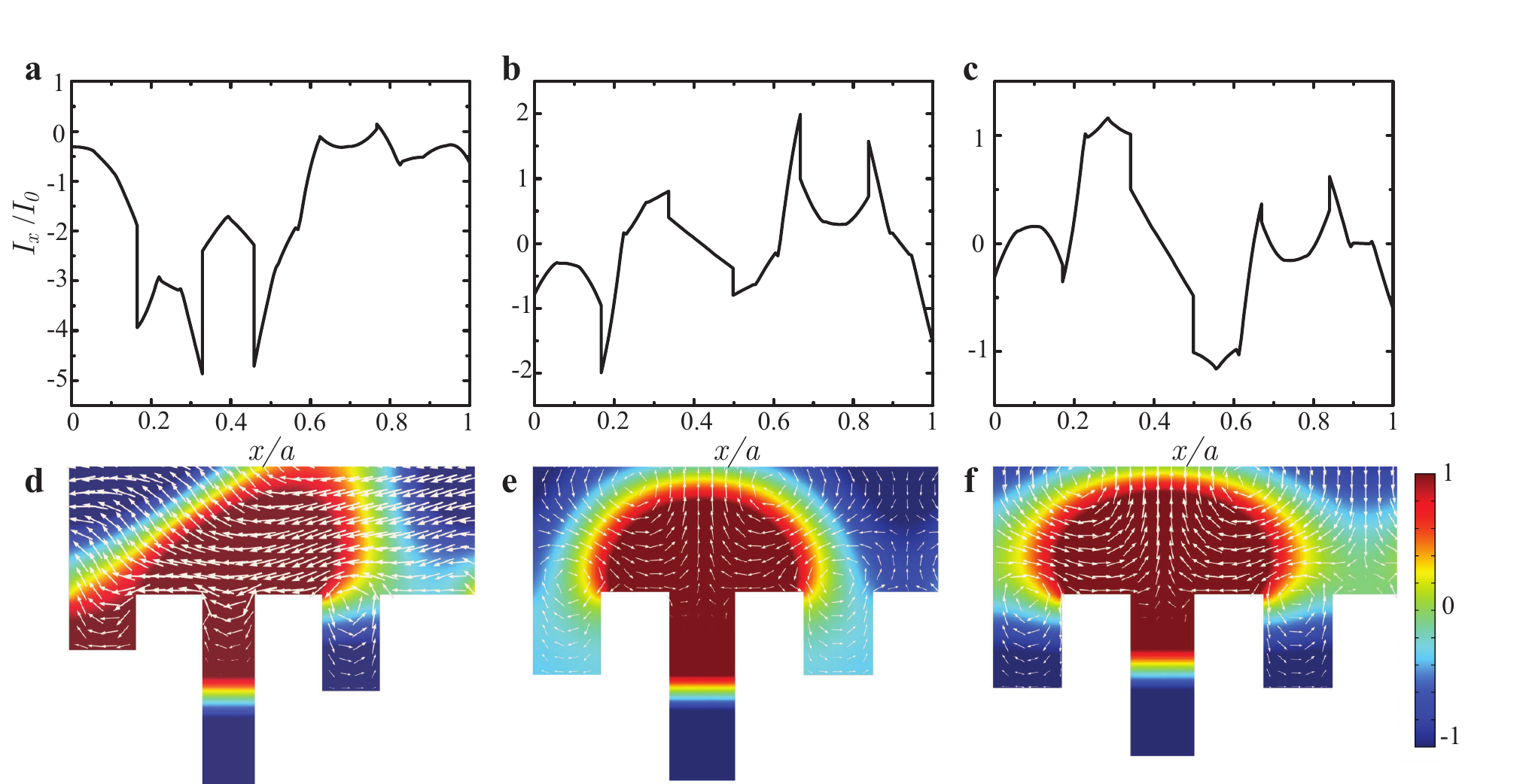}
\caption{\label{fig:3}  Intensity distribution in the total field. (a)-(c) Power flow value for ${{\bf{S}}_{\rm I}}$, ${{\bf{S}}_{{\rm I}{\rm I}}}$, and ${{\bf{S}}_{{\rm I}{\rm I}{\rm I}}}$ along $x$ direction at the surface of metasurface. The size and direction of white arrows in (d)-(f) respectively indicate the strength and direction of the power flow here.}
\end{figure}

As shown in Fig.~\ref{fig:1} and each matrix $\bf{S}$, along with these rectangular pits, which are attached randomly in one period, comes the non-local effect from metasurfaces. Compared with traditional gradient metasurface, additional degree of freedom for sound regulation could be utilized in these metasurfaces, which supports the lateral transport of energy flow. To demonstrate more details of non-local coupling, the energy flow at $y$ = 0 for these three metasurfaces in simulation has been extracted in Figs. \ref{fig:3}a-c. The local intensity vector ${I_x}$ could be obtained by ${I_x} = \frac{1}{2}\operatorname{Re} [p{({v_x})^ * }]$, where $p$ means pressure and the ${v_x}$ is the local velocity in the $x$ direction. The sign of ${{{I_x}} \mathord{\left/
 {\vphantom {{{I_x}} {{I_0}}}} \right.\kern-\nulldelimiterspace} {{I_0}}}$ refers to the direction of energy flow in $x$-axis. From Fig. \ref{fig:3}a, ${{{I_x}} \mathord{\left/
 {\vphantom {{{I_x}} {{I_0}}}} \right.
 \kern-\nulldelimiterspace} {{I_0}}}$ are almost negative, which suggests the sound energy is concentrated and reflected in the left channel. In contrast, the curves in Figs. \ref{fig:3}b-c float above and below zero, which indicates that the structures ${{\bf{S}}_{{\rm I}{\rm I}}}$, and ${{\bf{S}}_{{\rm I}{\rm I}{\rm I}}}$ not only allocate energy to the left channel but also allocate to the others. Figs. \ref{fig:3}d-f confirm this point more intuitively, corresponding to Figs. \ref{fig:3}a-c, respectively, which reveal the strong lateral energy exchange between grooves. In any case, metasurfaces via non-local effects realize efficient wave manipulation by using lateral energy transport in the form of etched grooves. Although the efficiency of non-local metasurfaces could be perfectly exerted in sound manipulation, the inverse calculation is complicated due to the randomness of its geometrical parameters. Herein, deep learning vastly simplifies it and facilitates the mass production for complex meta-structures. Not only that, it enables designers without theoretical foundations to quickly implement on-demand design. 
 
\begin{figure}
\includegraphics[width=15.6cm]{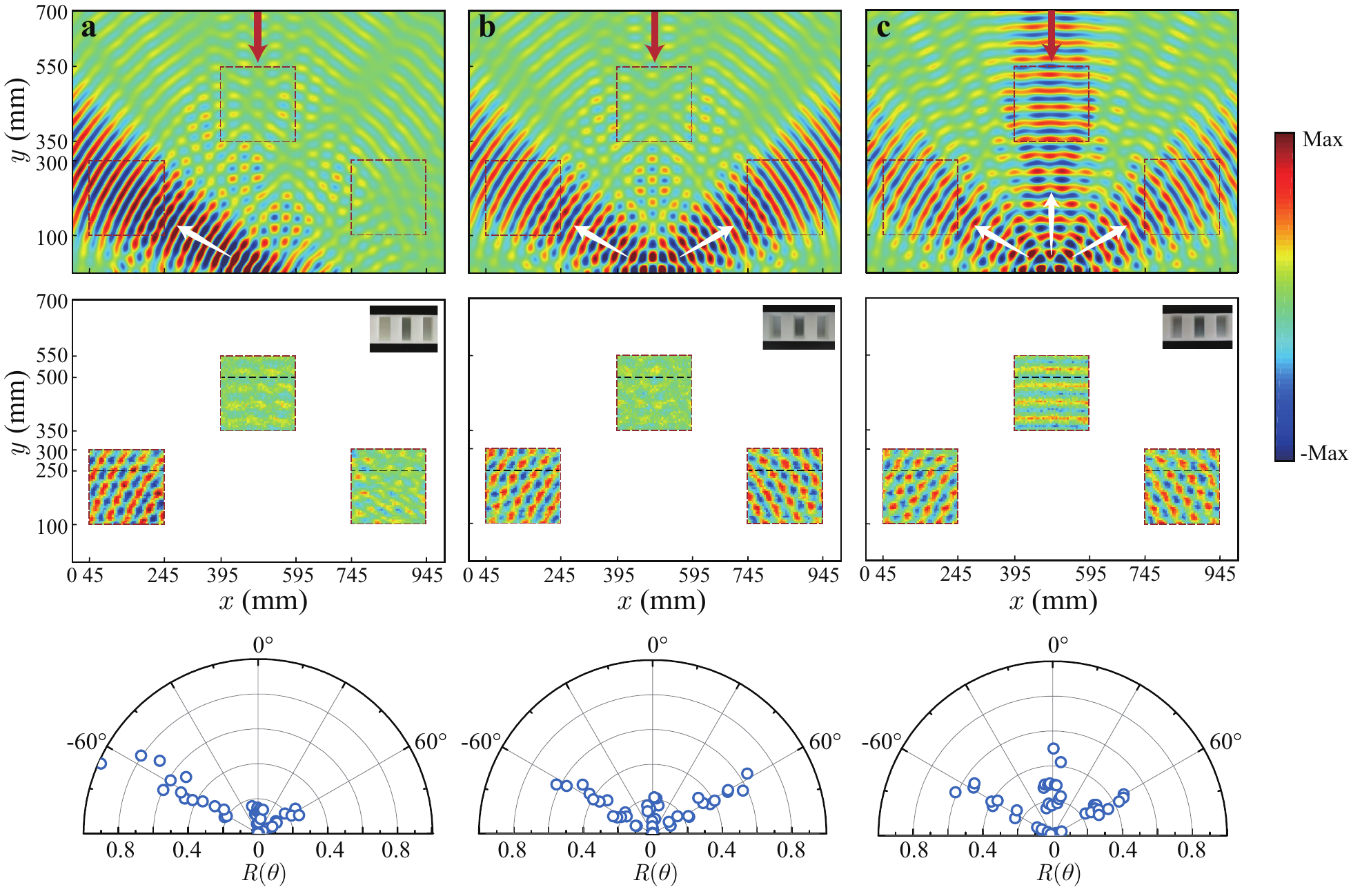}
\caption{\label{fig:4}  Finite element simulation and experimental results for scattered pressure fields. (a)-(c) Scattered fields corresponding to ${{\bf{S}}_{\rm I}}$, ${{\bf{S}}_{{\rm I}{\rm I}}}$,  ${{\bf{S}}_{{\rm I}{\rm I}{\rm I}}}$. The upper panels are the results of the simulation. To show the reflected fields more clearly, incident waves are not shown in the figures. Middle panels are experimental measurement fields, in which each measured area is a square with a size of $200 \times 200$ mm$^2$. The insets in middle panels are 3D-printed sample photos (one period is shown). The values at black dashed lines in the middle panels are extracted and plotted in the lower panels, where the abscissa $R$($\theta$) is the ratio of the real part for ${p_{\theta}}$ and ${p_{max}}$. ${p_{\theta}}$ represents the pressure value at different positions on the black dotted lines.}
\end{figure}

 \noindent \textbf{Experimental demostrations.} When a plane wave fed perpendicularly into the metagratings, scattering field with different ratios are represented in  Figs. \ref{fig:4}a-c, corresponding to ${{\bf{S}}_{\rm I}}$-${{\bf{S}}_{{\rm I}{\rm I}{\rm I}}}$ respectively. The red arrows in the figure indicate the position and direction of the incident wave whilst white arrows represent reflection. The reflected sound waves are obviously emitted from three channels we set, which follow the target energy distribution.
 
The experimental samples with 20 periods are fabricated by 3D-printing technology using the stereo lithography apparatus with photosensitive resin. Each measurement area is marked with a red dashed frame in the upper panels of Figs. \ref{fig:4}a-c and the size is $200 \times 200$ mm$^2$. The experimental results shown in the middle of Fig. \ref{fig:4} correspond to the scattering fields with target ratios (${{\bf{I}}_{\rm I}}$-${{\bf{I}}_{{\rm I}{\rm I}{\rm I}}}$). All the scattering fields in experiment have been normalized by the incident pressure. In addition, we select a line in each measurement area and plot the data extracted from the line in lower panels, defining $R(\theta ) = \operatorname{Re}({{{p_{\theta }}} \mathord{\left/
 {\vphantom {{{p_{\theta }}} {{p_{\max }}}}} \right.
 \kern-\nulldelimiterspace} {{p_{\max }}}})$
as the abscissa. Experimental results agree well with the simulation, which demonstrates the fantastic functionality of the metasurfaces designed by deep learning algorithm. In the last experimental field, the pressure field of the middle channel is weaker than the two others, which may be caused by the multiple reflection between loudspeaker array and the metasurface.
 
\noindent \textbf{Discussion} 

\noindent In this work, we investigate the beam splitting function of the grooved metasurfaces based on nonlocal coupling mechanism and realize the arbitrary regulation of the energy ratio for the multi-channel metasurfaces. We successfully conducted model training on $\bf{S}$ and $\bf{I}$ taking into account the nonlocality among subunits, so that finally the transient reactor can predict the target structures. Meanwhile, the efficiency and accuracy of the network model are validated by simulations and experiments. This intelligent method fairly shortens the optimization time for acoustic device design and this way could be extended to other forms of device structures. Our work provides new design solutions for non-local acoustic functional devices and promotes explorations of more fancy properties of meta-materials.These results might inspire more combinations of acoustics and artificial intelligence algorithms to assist in solving acoustic problems. 

\noindent\textbf{Method} \\
\noindent\textbf{Neural network training mechanism. }Given that one network model has $D$ layers, then the first layer gets the input data $X^{0}$ and the layer $D$ outputs the data $X^{D}$ which
has been activated many times by nonlinear activation functions. $f$ represents the activation function.The complete neural network training process is mainly composed of two parts. In the former process, neural network parameters $w$ and $b$ are used to calculate the output values in conjunction with the activation function. Forward equations are as follows:
\begin{equation}
   {{z^d} = {w^d}{X^{d - 1}} + {b^d}} \quad {for} \quad {d = 1,2, \cdots ,D}  \\ 
,
\label{eq:14}
\end{equation}
\begin{equation}
   {{X^d} = f({z^d})} \quad {for} \quad {d = 1,2, \cdots ,D}  \\ ,
\label{eq:15}
\end{equation}
where $z$ called weighted input will play an important role in the back propagation. Cost function($C$) evaluates the error between predicted values and expected values. Actually, the goal of training given data is to seek suitable model parameters for which cost function return as small value as possible. Whereas this purpose depends on the later process to complete. In the back propagation, parameters $w$ and $b$ could be updated by gradient calculation and the global optimizer to improve the predictive performance of models. Referring to the back propagation algorithm proposed in Ref.\cite{Rumelhart1986Learning}, error gradient of each layer can be expressed as:
\begin{equation}
{\delta ^D} = \frac{{\partial C}}{{\partial {X^D}}}f'({z^D}),
\label{eq:16}
\end{equation}
\begin{equation}
   {{\delta ^d} = \left[ {{{({w^{d + 1}})}^T}{\delta ^{d + 1}}} \right] \odot f'({z^d})} \quad {for} \quad {d = 1,2, \cdots ,D - } 1,
\label{eq:17}
\end{equation}
Then the partial differential of the cost function to the internal parameters of the network can be written as:
\begin{equation}
\frac{{\partial C}}{{\partial {w^d}}} = {a^{d - 1}}{\delta ^d},
\label{eq:18}
\end{equation}
\begin{equation}
\frac{{\partial C}}{{\partial {b^d}}} = {\delta ^d}.
\label{eq:19}
\end{equation}
Note that every neuron will go through these calculation cycles. Considering that the calculation relationship between two neurons located in adjacent layers is similar, for the sake of simplicity, we just use the superscript representing the number of layer for these parameters in above equations instead of marking the specific position of each neuron in a certain layer.

\noindent\textbf{Network model structure and hyperparameter settings. }In order to balance the training speed and prediction accuracy of the network model, we adopt the batch training method, that is, part of the data is sent for training each time. Batch size is the number of samples sent for each training. In all our training process, batch size is set to 256. We use the Leaky Relu as the activation function for each layer. The initial learning rates of forward training and tandem training are 0.0014 and 0.0006, respectively. The architecture of ${A_{net}}$ is selected as 8-512-512-256-256-128-3 and ${B_{net}}$ is 3-256-256-128-128-128-8. The 8 nodes match the number of structure parameters $\bf{S}$, and the 3 nodes correspond to the number of elements in the vector $\bf{I}$. Although pre-trained model and connected model could converge well after training over a thousand times, we terminated all the model training until epochs reached 5000 times to make the models more stable. 

\noindent\textbf{Finite element simulation.} COMSOL MULTIPHYSICS is used as a preliminary verification tool for preliminary simulation verification of the predicted results. The pressure acoustic module in the frequency domain is used to simulate the effect of the metasurface. A fixed-length plane incident wave as the sound source is incident on metasurfaces.

\noindent\textbf{Experimental setup.} The phase shift and amplitude of pressure fields are collected by a 1/8-inch microphone (Br{\"u}el $\& $ Kj{\ae}r type 2670) and the data-acquisition hardware (NI PXI-4461 in NI PXIe-1071). The microphone as a probe scans fields in a step of 5mm.  In the automatic scanning platform, we disposed 7 loudspeakers with a total width of 245 mm as the source. Samples are put in an organic glass plate waveguide about 2 cm high for experimental measurements. Foams are distributed at the edge of the measurement field to absorb external sound waves and reduce the influence of internal multiple reflection waves. We first measured the empty field with no sample placed, and then measured the total field with the placed sample. In this way, the difference between the two measurements is the result of the reflected field.

\noindent\textbf{Data availability} \\
\noindent The data that support the plots within this paper and other findings of this study are available from the corresponding author upon reasonable request.
%\bibliography{ref}
    % \printbibliography
%\bibliographystyle{naturemag}

\noindent \textbf{Acknowledgements} \\
\noindent This work is supported by the National Key R \& D Program of China under Grant Nos. 2020YFA0211400 and 2020YFA0211402, the National Natural Science Foundation of China under Grant No. 12074286, and the Shanghai Science and Technology Committee under Grant No. 20ZR1460900.

\noindent \textbf{Author contributions} \\
\noindent Y.L. conceived the idea. H.D. implemented model construction and training. X.F., and N.W. provides support for basic physical model. H.D., B.J., and N.W. performed the experiments. Y.L. and Q.C. supervised the whole project. H.D., X.F. and Y.L. wrote the paper with contribution from all authors.	

\noindent \textbf{Competing interests} \\
\noindent The authors declare no competing interests.
\end{document}